\documentclass{emulateapj}
\usepackage{color} 

\usepackage{epsfig}
\usepackage{subfigure}
\usepackage{graphicx}
\usepackage{amsmath}
\usepackage{natbib}
\usepackage{amssymb}
\usepackage{amsmath}
\usepackage{mathrsfs}

\newcommand{\mb}{\boldsymbol}

\shorttitle{Wind-driven Accretion in PPDs}
\shortauthors{X.-N. Bai}

\begin{document}


\title{Wind-driven Accretion in Protoplanetary Disks ---
II: Radial Dependence and Global Picture}


\author{Xue-Ning Bai\altaffilmark{1}}
\affil{Institute for Theory and Computation, Harvard-Smithsonian
Center for Astrophysics, 60 Garden St., MS-51, Cambridge, MA 02138}
\email{xbai@cfa.harvard.edu}

\altaffiltext{1}{Hubble Fellow}




\begin{abstract}
Non-ideal magnetohydrodynamical effects play a crucial role in determining the
mechanism and efficiency of angular momentum transport as well as the level of
turbulence in protoplanetary disks (PPDs), which are key to understanding PPD
evolution and planet formation. It was shown in our previous work that at 1 AU, the
magnetorotational instability (MRI) is completely suppressed when both Ohmic
resistivity and ambipolar diffusion (AD) are taken into account, resulting in a laminar
flow with accretion driven by magnetocentrifugal wind. In this work, we study the
radial dependence of the laminar wind solution using local shearing-box simulations.
Scaling relation on the angular momentum transport for the laminar wind is obtained,
and we find that the wind-driven accretion rate can be approximated as
$\dot{M}\approx0.91\times10^{-8}R_{\rm AU}^{1.21}(B_p/10{\rm mG})^{0.93}$
$M_{\bigodot}$ yr$^{-1}$,
where $B_p$ is the strength of the large-scale poloidal magnetic field threading
the disk. The result is independent of disk surface density.
Four criteria are outlined for the existence of the laminar wind solution: 1). Ohmic
resistivity dominated midplane region; 2). AD dominated disk upper layer; 3). Presence
of (not too weak) net vertical magnetic flux. 4). Sufficiently well ionized gas beyond disk
surface. All these criteria are likely to be met in the inner region of the disk from
$\sim0.3$ AU to about $5-10$ AU for typical PPD accretion rates.
Beyond this radius, angular momentum transport is likely to proceed due to a
combination of the MRI and disk wind, and eventually completely dominated by the MRI
(in the presence of strong AD) in the outer disk. Our simulation results provide key
ingredients for a new paradigm on the accretion processes in PPDs.
\end{abstract}


\keywords{accretion, accretion disks --- instabilities --- magnetohydrodynamics ---
methods: numerical --- planetary systems: protoplanetary disks --- turbulence}

\section{Introduction}\label{sec:intro}

Jets and molecular outflows are ubiquitous among young stellar objects (YSOs, see
\citealp{Ray_etal07} and \citealp{Arce_etal07} for a review). The velocity of the outflow
ranges from a few hundred km s$^{-1}$ for the axial high-velocity component (HVC)
that propagate to large distances, to about $10-50$ km s$^{-1}$ for the so-called
low-velocity component (LVC) that is more extended \citep{Hirth_etal97,Pyo_etal03}.
The outflow is strongly collimated at large distance ($\sim1000$ AU scale), with typical
opening angle of just a few degrees, while large opening angles ($20-30^{\circ}$) are
inferred at scales of 10 AU or less \citep{Woitas_etal02,Hartigan_etal04}. It has been
well established that outflow is associated with the accretion phenomenon, with the
mass outflow rate on the order of a tenth of the mass accretion rate
\citep{Cabrit_etal90,Hartigan_etal95,Cabrit07a}. The launching and collimation of the
jet/outflow in YSOs are almost certainly magnetic in nature \citep{Cabrit07}, and most
likely magneto-centrifugally driven via the \citet{BlandfordPayne82} mechanism:
outflowing gas launched from the disk can be accelerated along open magnetic field
lines centrifugally when the surface poloidal magnetic field is inclined by more than
$30^\circ$ relative to disk normal. The outflow can be gradually collimated beyond the
Alfv\'en point by the magnetic hoop stress from the toroidal field. The accretion-ejection
correlation leads to closer studies of the jet/outflow properties in the vicinity the
launching region. Using the Hubble Space Telescope, evidence of jet rotation has been
identified in a number of sources including DG Tau and RW Aur within the scale of about
100 AU from the star \citep{Bacciotti_etal02,Coffey_etal04,Coffey_etal07}. Although
caveats and alternative explanations do exist
\citep{Soker05,Cerqueira_etal06,Cabrit_etal06}, the jet rotation observations can be
naturally explained as originating from a magnetic outflow. Based on the conservation
laws for axisymmetric magnetic outflows, the location of the jet-launching
region can be inferred \citep{Anderson_etal03}, where it was found that the LVC wind
in these sources is likely to originate from an extended region in the disk from less than 1
AU to a few AU \citep{Anderson_etal03,Coffey_etal04,Coffey_etal07}. The launching
region of the HVC is not directly constrained due to the resolution limit, but is likely to be
located further inward, and may involve an X-wind \citep{Shu_etal94,Shang_etal07}.
Further investigations show that the LVC outflow in these systems carries a substantial
fraction (if not all) of the disk angular momentum in the launching region
\citep{Chrysostomou_etal08,Coffey_etal08}.
These observations reveal that the magnetocentrifugal wind is not only responsible for
producing the LVC of the outflow in YSOs, but is also likely to play a dominant role in
driving accretion for the inner region of protoplanetary disks (PPDs).

Typical accretion rate from PPDs in T-Tauri systems has been well-established to be
about $10^{-8\pm1}M_{\bigodot}$ yr$^{-1}$ \citep{Hartmann_etal98}. Accretion in PPDs
has conventionally been largely attributed to be due to the magnetorotational instability
(MRI, \citealp{BH91}). Due to the weak ionization level which introduces non-ideal
magnetohydrodynamical (MHD) effects, especially near the midplane region
of the inner disk, accretion in the inner region of PPDs was thought to be layered
\citep{Gammie96}, with the MRI-driven accretion proceeding in the better-ionized surface
layer termed as the active layer, and a more or less laminar midplane region termed as
the dead zone. However, this conventional picture takes into account only the effect of
Ohmic resistivity, while weakly ionized gas suffers from two other non-ideal MHD effects,
namely the Hall effect and ambipolar diffusion (AD). Assuming Hall effect has a limited
impact on the MRI as suggested by simulations (\citealp{SanoStone02a,SanoStone02b},
although the Hall-dominated regime was not covered in these simulations,
\citealp{WardleSalmeron12}), it was shown that the inclusion of AD dramatically reduces
the efficiency of the MRI, making it far insufficient to account for the large accretion
rate observed in typical T-Tauri disks
\citep{BaiStone11,Bai11a,Bai11b,PerezBeckerChiang11a,PerezBeckerChiang11b,Mohanty_etal13}.
These results strongly suggest magnetized outflow as a promising alternative to drive disk
accretion, in line with observations \citep{Bai11a}.

In our previous paper (\citealp{BaiStone13b}, hereafter paper I), we have performed a
series of local shearing-box simulations for a local patch of PPD at a fixed radius of
$R=1$ AU that for the first time, included both effects of Ohmic resistivity and AD, with
magnetic diffusion coefficients self-consistently determined from a pre-computed
look-up table in real simulation time. It was shown first that net vertical magnetic flux is
essential for driving rapid accretion (otherwise the MRI operates in an extremely inefficient
way due to AD), in which case even the initial field configuration is unstable to the MRI, the
system quickly relaxes to a state where the MRI is completely suppressed. This is due
to large Ohmic resistivity near the disk midplane, and strong AD in the low-density disk
surface where the magnetic field becomes too strong for the MRI to operate in the AD
dominated regime \citep{BaiStone11}. Instead of the MRI, the disk launches a strong
outflow that is magnetocentrifugal in nature. With a physical wind geometry (i.e., the
poloidal field/stream lines bend toward the same direction in the shearing-box), a very
weak vertical net flux with midplane plasma $\beta_0=10^5$ (ratio of gas pressure to
magnetic pressure of the net vertical field) is sufficient to drive rapid disk accretion with
accretion rate of a few times $10^{-8}M_{\bigodot}$ yr$^{-1}$. Moreover, accretion is
likely to proceed through a thin layer where the horizontal magnetic field changes sign
(strong current layer), while the rest of the disk remains static. This layer is offset from
disk midplane (at about 3 scale heights) and contains a tiny fraction of disk mass but
has large radial drift velocity (up to $0.4$ times the sound speed) to carry the accretion
mass flux.

The results from paper I is on the one hand surprising that MRI no longer operates
in the inner disk and the conventional picture of the MRI-driven layered accretion
becomes problematic since AD was not considered.
On the other hand, the results lend strong theoretical support to the observational
results of jet and outflows from YSOs summarized at the beginning of this section.
We note that most earlier theoretical work on disk wind from PPDs (e.g.,
\citealp{WardleKoenigl93,Li95,FP95,Konigl_etal10,Salmeron_etal11}) generally conclude
that strong net vertical magnetic field of equipartition level at disk midplane is needed
for wind launching. This is mostly due to the fact that a constant Elsasser number
(characterizing the coupling between gas and magnetic field) profile is assumed
across the vertical height of the disk. The results from paper I indicates
that with a realistic prescription on the vertical profile of the Ohmic and ambipolar
diffusion coefficients, launching of a laminar wind can be achieved with much weaker
vertical fields (in fact, equipartition-level magnetic field near the launching point,
which is high above the midplane), a result that was noticed \citep{Li96,Wardle97} but
not widely acknowledged.

In Paper I we have focused on the physical properties and the launching mechanism
of the newly-found laminar wind solution, obtained at fixed radius of 1 AU in a
minimum-mass solar nebular disk (MMSN, \citealp{Weidenschilling77}). It remains to
explore the radial range where such laminar wind solution exist. Toward smaller radii,
the disk becomes hotter and eventually thermal ionization of Alkali species takes
place (which essentially removes all non-ideal MHD effects).
The laminar wind picture would fail, and the gas dynamics is better described by MRI
in the ideal MHD regime.
Toward larger radii, the gas in PPDs gradually enters the AD dominated regime
due to the low density (beyond $\sim20$ AU). It has been predicted that MRI-driven
accretion can successfully account for the observed accretion rate
\citep{Bai11b,PerezBeckerChiang11b},
which was later confirmed in numerical simulations, where again, vertical net
magnetic flux turns out to be essential (\citealp{BaiStone11,Simon_etal13a,Simon_etal13b},
see Section \ref{ssec:global} for more discussions).

The main purpose of this paper is to identify the criteria (e.g., radial extent) where
the laminar wind solution exists, with main focus on the radial dependence of the
wind properties. Based on the discussion above, such laminar wind region must exist
in the vicinity of 1 AU, but does not extend to the inner edge of the disk, nor to the AD
dominated outer disk. We therefore perform the same simulations as in paper I (but
with some minor updates) at different disk radii from 0.3 AU to 15 AU and study the
wind properties, especially the role of net magnetic flux on wind-driven accretion, as
well as the transition from pure laminar wind toward the radius where MRI starts to set
in. Piecing together all the simulations results, a global picture concerning the gas
dynamics and evolution of PPDs is ready to be revealed.

This paper is organized as follows. We describe our numerical method, model
parameters and simulation diagnostics in Section \ref{sec:method}. Simulation results
in quasi-1D and 3D are presented in Sections \ref{sec:1dresults} and
\ref{sec:3dresults} respectively. In Section \ref{sec:discussion} we discuss the criteria for
the existence of laminar wind solution and propose a new global picture on the accretion
process in PPDs, together with concluding remarks.

\section[]{Simulations}\label{sec:method}

\subsection[]{Simulation Setup}

The methodology adopted in this paper follows exactly from paper I (with minor updates),
which we briefly summarize below. Readers should consult the entire Section 2 of
\citet{BaiStone13b} for details.

\begin{table}
\caption{List of All Simulation Runs.}\label{tab:runs}
\begin{center}
\begin{tabular}{ccccc}\hline\hline
 Run & $R$ (AU) &  $\beta_0$ & Dimension & Section\\\hline
S-R03-b5 & 0.3 & $10^5$ & quasi-1D & 3\\
S-R03-b6 & 0.3 & $10^6$ & quasi-1D & 3\\
S-R1-b4 & 1 & $10^4$ & quasi-1D & 3\\
S-R1-b5 & 1 & $10^5$ & quasi-1D & 3\\
S-R1-b6 & 1 & $10^6$ & quasi-1D & 3\\
S-R3-b4 & 3 & $10^4$ & quasi-1D & 3\\
S-R3-b5 & 3 & $10^5$ & quasi-1D & 3\\
S-R3-b6 & 3 & $10^6$ & quasi-1D & 3\\
S-R5-b4 & 5 & $10^4$ & quasi-1D & 3\\
S-R5-b5 & 5 & $10^5$ & quasi-1D & 3\\
S-R5-b6 & 5 & $10^6$ & quasi-1D & 3\\
S-R8-b4 & 8 & $10^4$ & quasi-1D & 3\\
S-R8-b5 & 8 & $10^5$ & quasi-1D & 3\\
F-R3-b5 & 3 & $10^5$ & 3D & 4\\
F-R3-b6 & 3 & $10^6$ & 3D & 4\\
F-R5-b5 & 5 & $10^5$ & 3D & 4\\
F-R5-b6 & 5 & $10^6$ & 3D & 4\\
F-R10-b4 & 10 & $10^4$ & 3D & 4\\
F-R10-b5 & 10 & $10^5$ & 3D & 4\\
F-R15-b4 & 15 & $10^4$ & 3D & 4\\
F-R15-b5 & 15 & $10^5$ & 3D & 4\\
\hline\hline
\end{tabular}
\end{center}
MMSN disk model, X-ray luminosity of $L_X=10^{30}$ ergs s$^{-1}$ and temperature
$T_X=5$keV, $0.1\mu$m grain with mass fraction of $10^{-4}$ are assumed for all runs.
\end{table}

We use the Athena MHD code \citep{Stone_etal08} and perform three-dimensional (3D)
shearing-box simulations for a local patch of a PPD at some fiducial radius $R$ with
Keplerian frequency $\Omega$. Radial, azimuthal and vertical dimensions are described
by $x$, $y$ and $z$ coordinates. The gas is assumed to be isothermal with
$P=\rho c_s^2$, where $\rho$, $P$ and $c_s$ are gas density, pressure and isothermal
sound speed respectively. Gas velocity with Keplerian velocity subtracted is denoted by
${\mb v}$. Magnetic field is denoted by ${\mb B}$, and we use the unit where magnetic
permeability is $1$ (so that magnetic pressure is simply $B^2/2$ and so on). Vertical
gravity from the central star is included, hence the simulations are vertically stratified, with
disk scale height $H\equiv c_s/\Omega$. In our code unit, we have
$\rho_0=\Omega=c_s=H=1$, where $\rho_0$ is the midplane gas density in hydrostatic
equilibrium, while physically we consider a MMSN disk at radius $R$. Unless otherwise
noted (only in one of the simulation runs), we use a density floor of $\rho_{\rm Floor}=10^{-6}$.

Ohmic resistivity and AD are included in the simulations, and we use $\eta_O$ and $\eta_A$
to denote Ohmic and ambipolar diffusivities. These diffusivities depends on the number
density of all charged species, with $\eta_O$ being independent of magnetic field strength,
while in general $\eta_A\propto B^2$ (with complications in the presence of small/tiny
grains, see Figure 1 of \citealp{Bai11b}). Their values are obtained by interpolating from a
pre-computed look-up table, given the local density and ionization rate in real simulation time.
The look-up table is computed by solving a complex chemical reaction network developed in
\citet{BaiGoodman09} and \citet{Bai11a} based on rate coefficients from the UMIST
database \citep{Woodall_etal07}. For chemical composition, we assume solar abundance
and $0.1\mu$m sized grains with mass fraction of $10^{-4}$ are included. We do not vary
grain abundance for all simulations in this paper, mainly because that it was found in paper
I that the properties of the wind is very insensitive to grain abundance (see their Figure 13
and explanations in section 5.3). The ionization rate
is obtained based on the horizontally averaged column density to the top and bottom of
the simulation box, using standard prescriptions for cosmic-ray and X-ray ionizations, as
well as ionization rate due to radioactive decay. Furthermore, we adopt a simplified
prescription to mimic the effect of FUV ionization based on the model by
\citet{PerezBeckerChiang11b} by assuming some fixed ionization fraction $f$ in the FUV
layer. The FUV layer lies in the disk surface within the column density of
$\Sigma_{\rm FUV}$ within which we use Equation (8) of paper I to compute magnetic
diffusivities, otherwise the diffusivities are interpolated from the look-up table.

In this work, we have slightly modified the prescriptions of the FUV ionization by
assuming two FUV ionization layers due to carbon and sulfur respectively, with a
sulfur ionization layer with $f=10^{-5}$ and $\Sigma_{\rm FUV}^{S}=0.03$ g cm$^{-2}$
and a carbon ionization layer that lies further above with $f=10^{-4}$ and
$\Sigma_{\rm FUV}^{C}=0.003$ g cm$^{-2}$. This prescription is, of course, still very
coarse, while it reflects the fact that the ionization fraction increases with height, and
alleviates the computational load by allowing larger time steps for calculations in the
outer disk (which is relevant only in our simulations with largest radius, i.e., $15$ AU).
We emphasize that the details in the prescription of the FUV ionization does
not affect the physical picture of MRI suppression and wind launching as discussed
in paper I. The role of FUV ionization is mainly to make the gas in the disk surface
layer well coupled to the magnetic field (i.e., approximately in the ideal MHD regime).


In this paper, we consider only two parameters, namely, the radial location in the disk
$R$ and the net vertical magnetic flux, characterized by
$\beta_0=2P_0/B_{z0}^2$, the ratio of midplane gas pressure ($P_0=\rho_0c_s^2$)
to the magnetic pressure of the net vertical field ($B_{z0}^2/2$). All other parameters
are fixed as described above. The role played by many of these other parameters,
such as grain abundance, the depth of the FUV ionization and disk surface density
are already explored in Section 5 of paper I, hence we do not repeat.

We conduct two types of simulation runs with different dimensionalities. The first
set of runs are fully three-dimensional (3D), with box size being $4H\times8H\times16H$
in $(x, y, z)$, resolved by $96\times96\times384$ cells. The box size is generally large
enough to fit potential MRI modes and these simulations are necessary to identify the
transition from fully laminar region close to 1 AU as studied in paper I to the MRI
dominated outer disk region. The second set of runs are quasi-one-dimensional
(quasi-1D) similar to those presented in Section 4 of paper I, where the vertical domain
size remains the same ($16H$) with slightly higher resolution (512 cells), while the
horizontal domain is shrinked to $4\times4$ cells (and cells have the same aspect ratio
as that in the first set of runs). These simulations will reproduce any 3D
simulations when the flow is completely laminar while is computationally much cheaper.
They are quasi-1D because a pure 1D run has difficulty to evolve into the final laminar
state from initial conditions (paper I). All simulation are initialized with a uniform vertical
magnetic field $B_{z0}$ characterized by $\beta_0$ and random velocity kicks on top
of the hydrostatic equilibrium configuration. In addition, a sinusoidally varying vertical
field with amplitude $B_{z1}=4B_{z0}$ is included to avoid the strong initial transient
channel flows. All simulations are run for $1200\Omega^{-1}$ ($200$ orbits, unless
otherwise noted) for adequate extraction of time-averaged quantities.

The list of all simulation runs conducted for this paper is shown in Table \ref{tab:runs}.
The range of radius $R$ is from 0.3 AU up to 15 AU. The quasi-1D simulations focus
on the inner region, from 0.3 AU to 8 AU  where the laminar wind solution is likely to
hold; while the full 3D runs focus on regions further out, from 3AU to 15 AU where MRI
may possibly set in. Since the inner boundary of this laminar zone depends on the
location where thermal ionization takes place, which requires the detailed study of disk
thermodynamics and is beyond the scope of this paper, we simply focus on the radial
dependence of laminar wind properties without asking the location of its inner laminar zone
boundary. On the other hand, the 3D runs that properly accommodate the MRI modes will
help identify outer boundary of the laminar wind zone.

\subsection[]{Diagnostics}\label{sec:diag}

Before we present simulation results, we describe the main diagnostics in our
simulations. Since all our simulations have net vertical magnetic flux, launching of disk
outflow is inevitable\footnote{The properties of disk wind also depend on global field
geometry, etc., but the launching of outflow is a local process enabled by the presence
of net vertical magnetic flux.}. In the case of laminar wind studied in detail in paper I
(which also applies to the quasi-1D simulations in our Section \ref{sec:1dresults}), we
have separated the system into a disk zone containing disk midplane and wind
zones at disk surface. The transition from disk zone to wind zone occurs when the
azimuthal gas velocity switches from sub-Keplerian to super-Keplerian (i.e., $v_y$
changes sign in our shearing-box simulations). This location is also referred to as
the base of the wind, beyond which the magnetocentrifugal mechanism takes place
and efficiently accelerates the outflow as it leaves the simulation domain. In 3D
simulations where MRI starts to set in but not yet dominant (Section \ref{sec:3dresults}),
one can still obtain time and horizontally averaged vertical profiles and define the
disk and wind zones based on the same criterion. With this definition, the main
diagnostics for our simulations include
\begin{itemize}
\item The $R\phi$ component of the stress tensor in the disk zone, which is
directly related to angular momentum transport in the radial direction. It can be
written as the sum of Reynolds (hydrodynamic) and Maxwell (magnetic)
components
\begin{equation}
T_{R\phi}\equiv T_{R\phi}^{\rm Rey} + T_{R\phi}^{\rm Max}
=\overline{\rho v_xv_y}-\overline{B_xB_y}\ ,
\end{equation}
where the overline denotes horizontal- (and time- except in Figure \ref{fig:fid3dhist})
average. Vertically integrating $T_{R\phi}$ across the disk zone, one obtains the
Shakura-Sunyaev $\alpha$ parameter
\citep{ShakuraSunyaev73}
\begin{equation}
\alpha\equiv\frac{\int_{-z_b}^{z_b}T_{R\phi}dz}{c_s^2\int \rho dz}\ ,\label{eq:alpha_def}
\end{equation}
where $z_b$ is the location of the base of the wind.
Similarly, we use $\alpha^{\rm Max}$ and $\alpha^{\rm Rey}$ to denote
contributions from the Maxwell and Reynolds components.

\item Location of the base of the wind $z_b$ (where $\bar{v}_y$ switches sign), as
well as the location of the Alfv\'en point $z_A$, defined as the point where the
gas vertical velocity equals the vertical Alfv\'en velocity
$\bar{v}_z=\sqrt{\overline{B_z^2}/\bar{\rho}}$. Note that in the presence of turbulence,
turbulent r.m.s. vertical field is adopted.

\item Rate of mass outflow from the top and bottom sides of disk surfaces
$\dot{M}_w\equiv|\overline{\rho v_z}|_{\rm top}+|\overline{\rho v_z}|_{\rm bot}$.
We note that in shearing-box simulations, the value of $\dot{M}_w$ is decreases with
increasing box height hence is not well constrained. On the other hand, its dependence
on various physical parameters remains meaningful (as discussed in Section 4.5 of paper I).

\item The $z\phi$ component of Maxwell stress at $z_b$, which is directly related
to angular momentum transport in the vertical direction. It is defined as
\begin{equation}
T_{z\phi}|_{z_b}=T_{z\phi}^{\rm Max}\equiv (-B_zB_y)|_{\pm z_b}\ .
\end{equation}
Note that the corresponding Reynolds stress is zero at $z_b$ by definition
($v_y|_{\pm z_b}=0$). As shown in paper I, $T_{z\phi}$ is roughly independent of
the simulation box height, hence it can be considered as a robust measure of the
wind-driven accretion rate.

\item The Elsasser numbers for Ohmic resistivity and AD, defined as
\begin{equation}
\Lambda\equiv\frac{v_A^2}{\eta_O\Omega}\ ,\qquad
Am\equiv\frac{v_A^2}{\eta_A\Omega}\ ,
\end{equation}
respectively, where $v_A^2=B^2/\rho$ is the Alfv\'en velocity. Note that
$\Lambda\propto B^2$, while $Am$ is largely independent of $B$ since in
general $\eta_A\propto B^2$. The Elsasser numbers characterize the relative
importance between the non-ideal MHD terms and the inductive term, and
non-ideal MHD terms dominate when $\Lambda, Am <1$. In particular, the
MRI will be suppressed when $\Lambda<1$, and when magnetic field is too
strong at a given $Am$ (see Figure 16 of \citealp{BaiStone11}). Although the
Hall term is not included in our simulations, we do compute the Hall Elsasser
number $Ha\equiv v_A^2/\eta_H\Omega$ for discussion purposes.
\end{itemize}

Assuming steady state accretion, the accretion rate can be estimated as
\begin{equation}
\begin{split}
\dot{M}=&\frac{2\pi}{\Omega}\int_{-z_b}^{z_b} dzT_{R\phi}
+\frac{4\pi}{\Omega}RT_{z\phi}\bigg|_{-z_b}^{z_b}\\
=&\frac{2\pi}{\Omega}\alpha c_s^2\Sigma
+\frac{8\pi}{\Omega}R|T_{z\phi}|_{z_b}\ ,\label{eq:mdot}
\end{split}
\end{equation}
where the two terms correspond to radial transport and vertical transport
respectively, and it is assumed that the outflow has a physical geometry
so that $T_{z\phi}|_{z_b}=-T_{z\phi}|_{-z_b}$. Plugging in the numbers for
a MMSN disk, we further have
\begin{equation}
\dot{M}_{-8}\approx0.82\bigg(\frac{\alpha}{10^{-3}}\bigg)R_{\rm AU}^{-1/2}
+4.1\bigg(\frac{|T_{z\phi}|_{z_b}}{10^{-4}\rho_0c_s^2}\bigg)R_{\rm AU}^{-3/4}\ ,\label{eq:accrete}
\end{equation}
where $\dot{M}_{-8}$ is the accretion rate normalized to $10^{-8}M_{\bigodot}$
yr$^{-1}$.

In many 3D simulations, the systems become prone to the MRI.
To quantify the effectiveness of turbulence, we further evaluate the $R\phi$
component of the turbulent Maxwell and Reynolds stresses
in all 3D runs, which are obtained by subtracting contributions due to large-scale
magnetic field and velocity field
\begin{equation}
\begin{split}
T_{R\phi, {\rm turb}}&\equiv T_{R\phi, {\rm turb}}^{\rm Rey} + T_{R\phi, {\rm turb}}^{\rm Max}\\
&=(\overline{\rho v_xv_y}-\bar{\rho}\bar{v}_x\bar{v}_y)-(\overline{B_xB_y}-\bar{B}_x\bar{B}_y)\ .
\end{split}
\end{equation}
Correspondingly, one can also define $\alpha_{\rm turb}$ for Maxwell and Reynolds
components following equation (\ref{eq:alpha_def}).

\subsubsection[]{Symmetry and Strong Current Layer}

Symmetry of the solutions has been discussed extensively in Section 4.4 of paper I. Here
we briefly summarize the main results. Since curvature is ignored in shearing-box
approximation, there is no distinction between radially inward and outward directions. A
physical wind picture requires that the outflows from the top and bottom sides of the box
bend toward the same direction, or equivalently, the radial and toroidal components of the
magnetic field bend toward the same direction. For the quasi-1D solutions obtained in
paper I (at 1 AU), it was shown that the bending directions at the top and bottom sides are
random, and they are independent, hence there is equal chance to obtain physical and
unphysical solutions. The unphysical solutions are smooth, and all three component of the
magnetic field remain the same sign throughout the disk, which is said to obey odd-$z$
symmetry. Toroidal field is the dominant component within the disk. The physical wind
solution requires the horizontal field to change sign across the disk. If this occurs at the
midplane, the solution would obey even-$z$ symmetry. However, simulations show that
the flip occurs in a thin layer that is offset from the midplane, forming a strong current layer.
This layer carries the entire accretion flow, while other than the flip of horizontal magnetic
field, the properties of the outflow are exactly the same as those in the solution
with odd-$z$ symmetry.

In paper I, it was also noted that the strong current layer does not seem to be long-lived
in 3D shearing-box simulations (see footnote 12 of paper I), and the odd-$z$ symmetry
solution (unphysical) is preferred.
This fact is likely due to the limitations of the shearing-box approximation, where curvature
is ignored and the vertical domain size is limited\footnote{Curvature may make the
physical wind geometry solution more preferable. Also, disk outflow from real disks is
likely weaker than obtained from typical shearing-box simulations due to limited box
height (paper I), which may make the strong current layer less susceptible to instabilities.}.
Recent 2D global disk simulations at least suggest the possible existence of strong current
layer offset from the disk midplane (e.g., the kink feature in Figure 13 of
\citealp{FendtShe13}, although their simulations involve much stronger magnetic field and
rough treatment of microphysics). Still, global simulations in 3D are essential to properly
study the behavior of the strong current layer. 

Since the properties of the outflow is largely independent of the symmetry of
the solution in shearing-box, in this work, we mainly focus on the outflow/wind properties.
In particular, for 3D simulations, the large-scale field of all runs end up with odd-$z$
symmetry, hence we do not label their symmetries separately. For quasi-1D simulations,
we report the symmetry of the solution (which is random) as ``odd" (unphysical) or ``even"
(physical, despite that the strong current layer is offset from the midplane). For physical
solutions, we further characterize the strong current layer by
\begin{itemize}
\item Maximum radial velocity $v^{\rm SC}_{R,{\rm max}}$ in the strong current layer.
We further use $z^{\rm SC}$ to denote the location where $v^{\rm SC}_{R,{\rm max}}$ is
achieved.

\item The thickness of the strong current layer $h^{\rm SC}$, estimated by
$h^{\rm SC}$ by
\begin{equation}
h^{\rm SC}\equiv\frac{\int_{\rm SC}\rho(z)v_x(z)dz}{\rho(z^{\rm SC})v^{\rm SC}_{R, {\rm max}}}\ ,
\end{equation}
where the integral is performed through the strong current layer, in which we use
$v^{\rm SC}_{R,{\rm max}}$ as a characteristic radial velocity.
\end{itemize}

\begin{table*}
\caption{Results for All quasi-1D Simulations with a Laminar Wind.}\label{tab:1dresults}
\begin{center}
\begin{tabular}{cccccccccc}\hline\hline
 Run & $\alpha^{\rm Max}$ & $T_{z\phi}^{\rm Max}$ & $\dot{M}_w$
 & $z_b$ & $z_{\rm A}$ & Symmetry & $|z^{\rm SC}|$ & $h^{\rm SC}$
 & $v^{\rm SC}_{R, {\rm max}}$ \\\hline

S-R03-b5 & $2.71\times10^{-4}$ & $7.17\times10^{-5}$ & $1.36\times10^{-5}$
               & $4.51$ & $6.92$ & Odd & - & - & - \\
S-R03-b6 & $4.10\times10^{-5}$ & $1.36\times10^{-5}$ & $4.67\times10^{-6}$
               & $5.05$ & $5.98$ & Odd  & - & - & -\\
S-R1-b4 & $1.26\times10^{-3}$ & $5.87\times10^{-4}$ & $8.82\times10^{-5}$
               & $3.92$ & $7.39$ & Odd & - & - & - \\
S-R1-b5 & $2.37\times10^{-4}$ & $1.04\times10^{-4}$ & $2.92\times10^{-5}$
               & $4.61$ & $6.13$ & Even  & $3.05$ & $0.13$ & $0.47$ \\
S-R1-b6a & $2.97\times10^{-5}$ & $2.73\times10^{-5}$ & $1.08\times10^{-5}$
               & $4.52$ & $5.36$ & Odd  & - & - & - \\
S-R1-b6b & $2.88\times10^{-5}$ & $2.42\times10^{-5}$ & $8.95\times10^{-6}$
               & $4.50$ & $5.50$ & Even  & $3.01$ & $0.35$ & $3.5\times10^{-2}$ \\
S-R3-b4 & $1.52\times10^{-3}$ & $7.92\times10^{-4}$ & $1.86\times10^{-4}$
               & $4.28$ & $6.33$ & Even  & $1.98$ & $0.17$ & $0.19$ \\
S-R3-b5a & $2.02\times10^{-4}$ & $2.01\times10^{-4}$ & $6.98\times10^{-5}$
               & $4.17$ & $5.27$ & Odd  & - & - & - \\
S-R3-b5b & $1.66\times10^{-4}$ & $1.96\times10^{-4}$ & $6.54\times10^{-5}$
               & $4.11$ & $5.33$ & Even  & $1.73$ & $0.80$ & $6.6\times10^{-3}$ \\
S-R5-b4 & $1.51\times10^{-3}$ & $1.06\times10^{-3}$ & $2.83\times10^{-4}$
               & $4.03$ & $5.80$ & Even  & $1.64$ & $0.23$ & $0.11$ \\
S-R5-b5 & $1.81\times10^{-4}$ & $2.53\times10^{-4}$ & $8.98\times10^{-5}$
               & $3.92$ & $5.08$ & Even  & $1.27$ & $1.25$ & $2.7\times10^{-3}$ \\
S-R8-b4 & $1.75\times10^{-3}$ & $1.40\times10^{-3}$ & $4.11\times10^{-4}$
               & $3.80$ & $5.33$ & Even  & $1.64$ & $0.23$ & $0.13$ \\
\hline\hline
\end{tabular}
\end{center}
Note that all numbers above are in code unit ($\rho_0=c_s=\Omega=1$). We
have conducted runs S-R1-b6 and S-R3-b5 twice labeled with ``a" and ``b"
respectively with different initial random seeds, and their resulting solutions end up
with different symmetries. Runs S-R3-b6, S-R5-b6 and S-R8-b5 turn out to be
unsteady, hence are not included in this table.
\end{table*}

\section[]{Results: Quasi-1D Simulations}\label{sec:1dresults}

Most of the quasi-1D simulations settle into pure laminar state with the structure of the
solution similar to that described in paper I. Major diagnostics are calculated when full
laminar states are reached for each of these runs with results listed in Table
\ref{tab:1dresults}. In order to study the properties of the strong current layer, for some
parameter sets, we conduct the same simulation for more than once with different
random seeds to find solutions with both symmetries. They apply to runs S-R1-b6 and
S-R3-b5, where we further use ``a" and ``b" to label solutions with odd-$z$ and even-$z$
symmetries respectively. In addition, we find that the gas density in run S-R03-b6 at
vertical boundaries hits the floor value of $10^{-6}$, hence we reduce floor value to
$10^{-7}$ in this particular run. 

Due to a slightly different treatment of the FUV ionization, we first calibrate our runs
S-R1-b$x$ with their counterparts in paper I (S-OA-b$x$), where $x=4, 5$ or $6$.
Comparing our Table \ref{tab:1dresults} with Table 4 of paper I, we see that all major
diagnostic quantities between the two sets of simulations agree very well (mostly within
$5\%$). This is not surprising since the adopted penetration depth of the FUV ionization
are the same ($0.03$g cm$^{-2}$), and the gas behaves closely to ideal MHD within
the FUV layer. In general, the introduction of the second the FUV ionization layer on top
of the first one has no influence on the properties of the solution for all our quasi-1D runs.

\subsection[]{A Fiducial Example}

\begin{figure*}
    \centering
    \includegraphics[width=170mm]{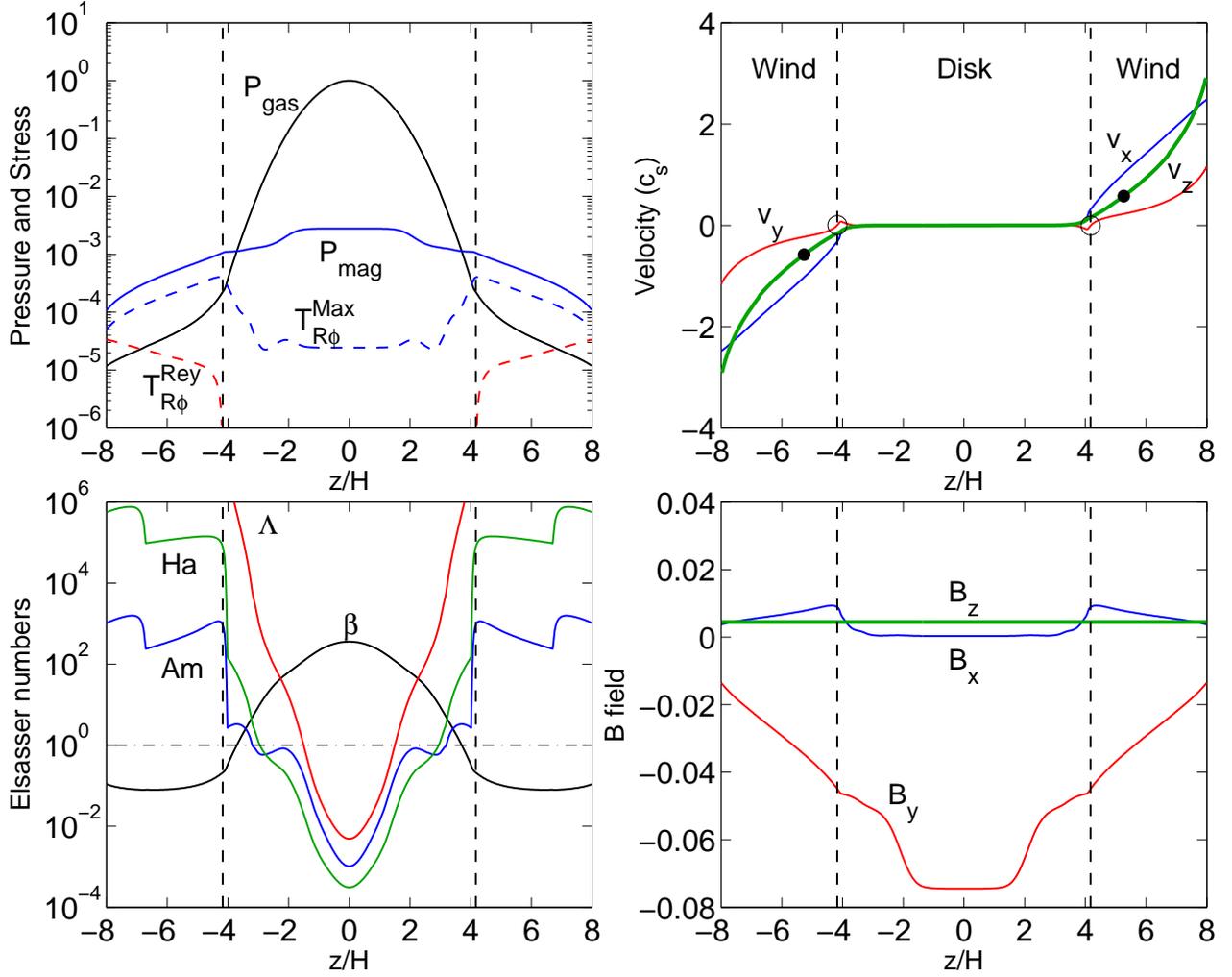}
  \caption{Vertical profiles of various quantities in the representative quasi-1D run
  S-R3-b5 at 3 AU with $\beta_0=10^5$ (with odd-$z$ symmetry).
  Upper left: gas pressure and magnetic pressure, as well as the $R\phi$ component of
  Maxwell and Reynolds stresses. Lower left: Elsasser numbers for Ohmic resistivity
  ($\Lambda$), Hall term ($Ha$) and AD ($Am$), together with
  $\beta=P_{\rm gas}/P_{\rm mag}$. Note that we evaluate $Ha$ for discussion purposes
  while the Hall effect is not included in the simulations. Upper right: three components of
  gas velocity, where the bold green curve is for vertical velocity. The Alfv\'en points
  are indicated as black dots, while open circles mark the base of the outflow. Lower right:
  three components of the magnetic field, where the bold green curve is for vertical field.
  In all panels, the vertical black dashed lines divide the domain into the disk zone (in the
  middle) and wind zones at the location of the wind base.}\label{fig:fid1d}
\end{figure*}

We pick run S-R3-b5 as a fiducial example to illustrate the laminar wind solution at 3 AU
in Figure \ref{fig:fid1d}. Here we pick the run with odd-$z$ symmetry (no strong current
layer) to be directly compared with its 1 AU counterpart shown in Figure 5 of paper I.
The basic properties of between the two solutions are very similar. The disk
zone is generally gas pressure dominated, and the magnetic field is dominantly toroidal.
The toroidal field profile near the midplane is almost flat, which is because of the
excessively large magnetic diffusion that makes the gas and magnetic field effectively
decoupled (hence current is not permitted). The vertical gradient of the toroidal magnetic
field gives radial current, which generates a Lorentz force in the azimuthal direction.
The balance of this force with Coriolis force leads to radial motion in the gas, which bends
the poloidal magnetic field. Upon achieving a sufficiently large bending angle ($>30^\circ$),
the wind is effectively launched and accelerated by the magnetocentrifugal mechanism.

Comparison between our Figure \ref{fig:fid1d} with Figure 5 of paper I reveal a few
trends about the dependence of the solution on disk radii. First of all, the magnetic
diffusion coefficients at disk midplane is substantially reduced at 3 AU relative to
those at 1 AU. In particular, the Ohmic resistivity cap and the floor on $Am$ are no
longer reached (see the end of Section 2 of paper I). Correspondingly, the
magnetically decoupled region (where field lines are straight) shrinks to
within $z=\pm2H$. This means that wind launching process takes place further
deeper toward disk midplane. Moreover, AD becomes more dominant at 3 AU. In
fact, if we do not consider the Hall effect, then the entire disk is AD dominated (this
is partly because magnetic field is strongest at disk midplane). Nevertheless, the
Ohmic Elsasser number is still well below $1$ at disk midplane, too small for MRI
to take place.

Secondly, the FUV ionization, assumed to have a fixed penetration depth in column
density, penetrates deeper into the disk, with the FUV ionization front located at
about $\pm4H$ instead of $\pm4.5H$. As discussed in paper I, deeper penetration
depth leads to larger wind mass loss rate (in code unit), as seen from Table
\ref{tab:1dresults}. Similarly, the
base of the wind $z_b$ and the Alfv\'en point are also located deeper at 3 AU. For
all our quasi-1D runs at the inner disk, the addition of the second FUV ionization
layers further higher above does not have any impact on the wind solution.

Together, with weaker magnetic diffusion and deeper FUV penetration, we see from
Table \ref{tab:1dresults} that the wind at 3 AU is about two times stronger (in
code/natural unit) than that at 1 AU in terms of the outflow rate $\dot{M}_w$, as well
as the wind stress $T_{z\phi}^{\rm Max}$, assuming that the background plasma
$\beta_0$ to be the same.

\subsection[]{Parameter Dependence}\label{ssec:1dresultsparam}

\begin{figure*}
    \centering
    \includegraphics[width=170mm]{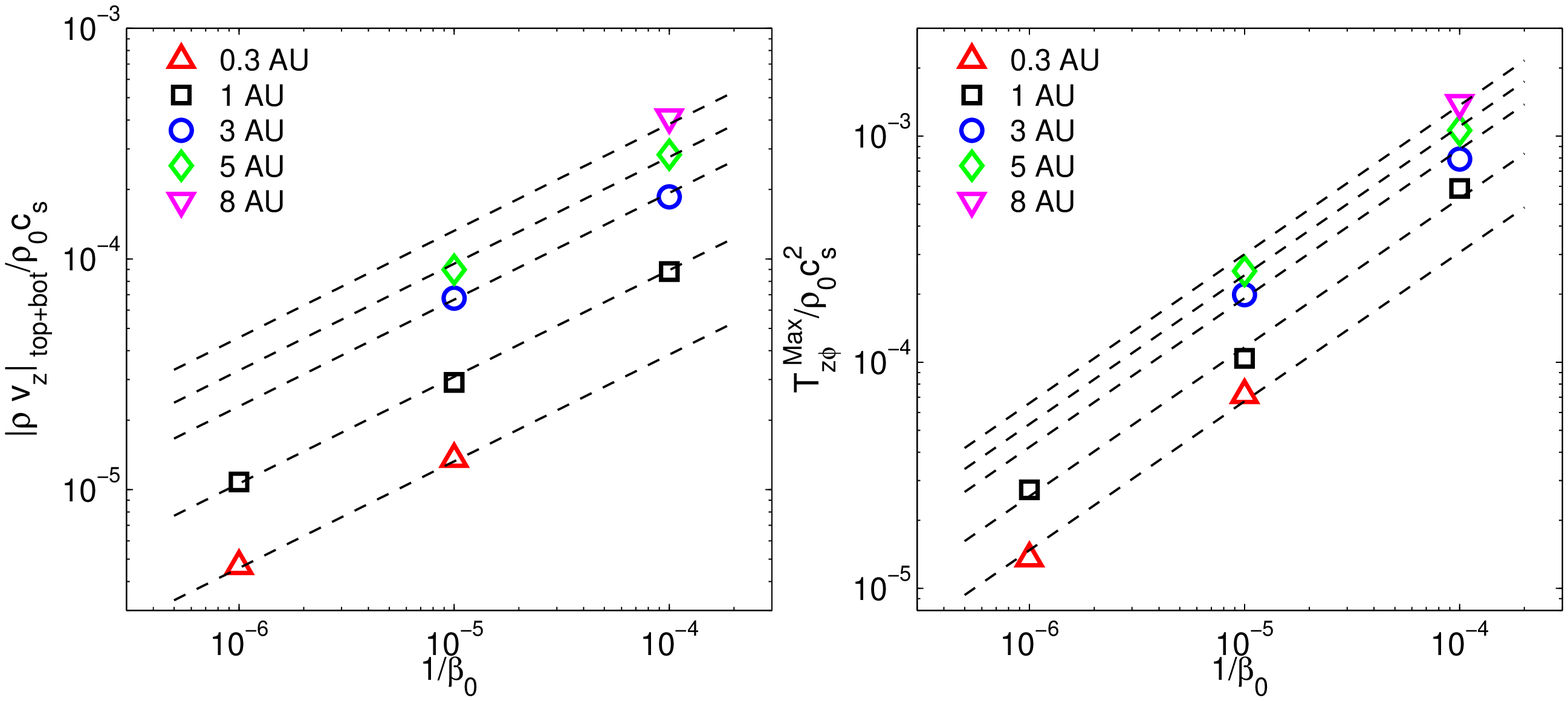}
  \caption{The wind mass loss rate $\dot{M}_w$ (left) and wind stress
  $T_{z\phi}^{\rm Max}$ (right) as a function of net vertical magnetic flux
  (characterized by $\beta_0$) measured from all our quasi-1D simulations with a
  laminar wind. Different symbols correspond to simulations at different radii, and
  dashed lines are fitting results for each radii, using equations (\ref{eq:fitmw})
  and (\ref{eq:fitTm}). All quantities are in code units, while see discussions in
  Section 3.2 for caveats.}\label{fig:1dparam}
\end{figure*}

In this subsection we discuss how the wind mass loss rate $\dot{M}_w$ and the
wind-driven accretion rate (due to $T_{z\phi}^{\rm Max}$) depend on net vertical
magnetic flux and disk radius.
In paper I, we found that at 1 AU, the wind stress $T_{z\phi}$ scales with
net vertical flux approximately as $\beta_0^{-0.7}$, and $\dot{M}_w$ scales with
net flux approximately as $\beta_0^{-0.5}$. With our new set of
simulations at different radii, we update their results with new simulation data
shown in Table \ref{tab:1dresults} and plotted on Figure \ref{fig:1dparam}.

Clearly, we see that the power-law dependence of $\dot{M}_w$ and
$T_{z\phi}^{\rm Max}$ on the net vertical magnetic flux persists at all radii as long
as the laminar wind solution exists. Assuming further a power-law dependence
on disk radii in the form of $CR^q\beta_0^{-b}$, where $C$ is a constant, we
find the following fitting results using linear regression
\begin{equation}\label{eq:fitmw}
\frac{\dot{M}_w}{\rho_0c_s}\approx3.08\times10^{-5}
\bigg(\frac{\Sigma_{\rm MMSN}}{\Sigma}\bigg)
\bigg(\frac{R}{\rm AU}\bigg)^{0.70}\bigg(\frac{\beta_0}{10^5}\bigg)^{-0.46}\ ,
\end{equation}
\begin{equation}\label{eq:fitTm}
\frac{T_{z\phi}^{\rm Max}}{\rho_0c_s^2}\approx1.16\times10^{-4}
\bigg(\frac{\Sigma_{\rm MMSN}}{\Sigma}\bigg)
\bigg(\frac{R}{\rm AU}\bigg)^{0.46}\bigg(\frac{\beta_0}{10^5}\bigg)^{-0.66}\ .
\end{equation}
Here we have further included a factor $\Sigma_{\rm MMSN}/\Sigma$, the inverse
of disk surface density normalized to the MMSN value (for reasons to be discussed).
The fitting results are also plotted in Figure \ref{fig:1dparam}, which we see that the
above fitting formulas match our simulation results perfectly.

We note that these fitting formulas are derived from local shearing-box simulations
of a MMSN disk. We recall from paper I (see their Section 5.4) that the wind mass loss
rate $\dot{M}_w$ and wind stress $T_{z\phi}^{\rm Max}$ in {\it physical unit} depend
only on the {\it physical strength} of the net vertical field. Therefore, to apply the formulas
to disks with different surface densities, two measures should be taken:
1). The factor of $\Sigma_{\rm MMSN}/\Sigma$ is included on the right hand side of the
formulas because the normalization factors on the denominator of the left hand side
contain $\rho_0\propto\Sigma$.
2). One must interpret $\beta_0$ as the ratio of midplane gas pressure {\it
of a MMSN disk} to the magnetic pressure of the net vertical field. In other words, for
a hypothetical disk with $\Sigma_1=N\Sigma_{\rm MMSN}$ and vertical plasma
$\beta_1$, one should set $\Sigma=\Sigma_1$ and $\beta_0=\beta_1/N$ in the formula
above.

For wind stress $T_{z\phi}^{\rm Max}$, it was found in paper I that its value is very
insensitive to vertical box height. Therefore, the value obtained from shearing-box
simulations provides a robust estimate of wind-driven accretion rate from Equation
(\ref{eq:mdot}).

The wind mass loss rate $\dot{M}_w$, unfortunately, is not well determined determined
from shearing-box simulations. As discussed in paper I, $\dot{M}_w$ roughly depends
inversely on the vertical size of the simulation domain $L_z$, and a physical value might
be achieved with $L_z\sim2R$. Since $H/R\ll1$ in PPDs, our Equation (\ref{eq:fitmw})
tend to significantly overestimate the wind mass loss rate in real disks. A possible
estimate of the true wind mass loss rate may be obtained by multiplying a factor of
$8H/R$ on the right hand side of Equation (\ref{eq:fitmw}), since the box height in all
our simulations is $L_z=16H$. It would be straightforward to apply the value of $H/R$
from MMSN or other disk models into the formula, but we do not proceed further here
since such estimate is very speculative. Future work with global simulations are
needed to provide a definitive answer.

The dimensionless form of the fitting formulas above have quantified the radial
dependence of wind properties. The indices on $\beta_0$ is slightly different but
consistent with those reported in paper I. In particular, they confirm the trend
discussed in the previous subsection that relatively stronger wind (in code unit) is
launched at larger radii due to weaker magnetic diffusion and deeper FUV penetration.

For a MMSN disk, we note that the index on $R$ in the fitting formula
(\ref{eq:fitTm}) is smaller than $3/4$. Therefore, using Equation (\ref{eq:accrete}),
we see that to maintain steady-state wind-driven accretion, the value of $\beta_0$
needs to decrease with radius as $\beta_0\propto R^{-(0.75-0.46)/0.66}=R^{-0.44}$
for a MMSN disk.
This is in accordance with our choice of smaller $\beta_0$ at larger disk radii.
We will further discuss the wind-driven accretion rate based on the fitting
formulas above in Section \ref{ssec:accretion}.

\subsection[]{The Strong Current Layer}

\begin{figure}
    \centering
    \includegraphics[width=85mm]{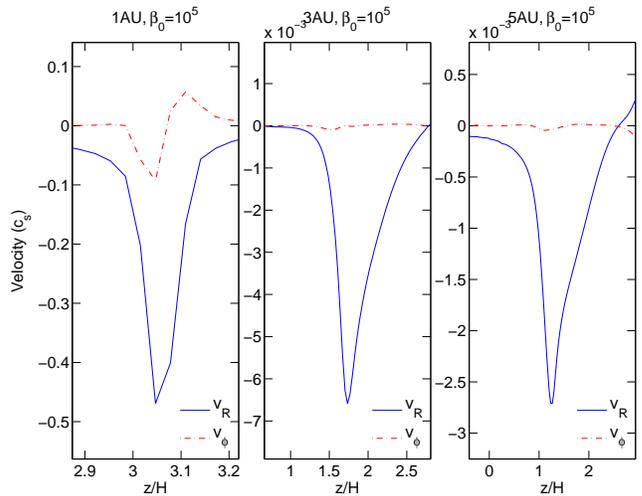}
  \caption{The radial (blue solid) and azimuthal (red dash-dotted) velocity profiles
  across the strong current layer in three quasi-1D simulation runs at 1 AU, 3 AU
  and 5 AU respectively, with fixed $\beta_0=10^5$. Note that we have rescaled
  the axis limits so that peak velocities are centered, and the size of each box is
  $4h^{SC}\times1.5v_{R, {\rm max}}^{SC}$.}\label{fig:scl}
\end{figure}

The properties of the strong current layer for most of our quasi-1D runs are given
in the last three columns of Table \ref{tab:1dresults}. We see two very clear trends:
1). Toward outer disk radii, the strong current layer shifts toward disk midplane,
becomes wider, with slower gas inflow; 2). With stronger net vertical magnetic
flux, the strong current layer shifts toward disk surface, becomes sharper, with
much faster gas inflow. As discussed in paper I, the strong current layer is offset
from the midplane since the midplane region is too poorly coupled to the gas
hence current is not permitted. It is generally located at the height where the gas
starts to be partially coupled with the magnetic field. Since the gas near midplane
region becomes progressively better coupled with magnetic fields, it is therefore
understandable that the location of the strong current layer shifts toward disk
midplane at larger disk radii. Also, stronger net vertical magnetic flux generates
stronger toroidal magnetic field that is more difficult to flip, which favors
better-coupled region (toward disk surface) to take place.

\begin{table*}
\caption{Summary of All 3D Simulations.}\label{tab:3dresults}
\begin{center}
\begin{tabular}{ccccccccc}\hline\hline
 Run & $\alpha^{\rm Max}$ & $\alpha^{\rm Rey}$ & $T_{z\phi}^{\rm Max}$ & $\dot{M}_w$
 & $z_b$ & $z_{\rm A}$
 & $\alpha_{\rm turb}^{\rm Max}$ & $\alpha_{\rm turb}^{\rm Rey}$ \\\hline 

F-R3-b5 &$1.96\times10^{-4}$ & $8.16\times10^{-6}$ & $2.02\times10^{-4}$ & $6.95\times10^{-5}$
		   & $4.15$ & $5.27$ & $2.10\times10^{-7}$ & $2.80\times10^{-6}$ \\
F-R3-b6 & $1.96\times10^{-4}$ & $1.47\times10^{-5}$ & $2.89\times10^{-5}$ & $8.03\times10^{-6}$
		   & $6.06$ & $-$ & $1.02\times10^{-6}$ & $4.52\times10^{-6}$ \\
F-R5-b5 & $2.07\times10^{-4}$ & $0.96\times10^{-6}$ & $2.67\times10^{-4}$ & $9.83\times10^{-5}$
		   & $3.97$ & $5.02$ & $5.16\times10^{-8}$ & $4.86\times10^{-7}$ \\
F-R5-b6 & $2.32\times10^{-4}$ & $2.32\times10^{-5}$ & $3.40\times10^{-5}$ & $6.33\times10^{-6}$
		   & $5.96$ & $-$ & $4.34\times10^{-6}$ & $7.56\times10^{-6}$ \\
F-R10-b4 & $2.53\times10^{-3}$ & $6.18\times10^{-5}$ & $2.16\times10^{-3}$ & $6.85\times10^{-4}$
		   & $3.44$ & $5.77$ & $-1.24\times10^{-4}$ & $2.21\times10^{-5}$ \\
F-R10-b5 & $2.40\times10^{-3}$ & $2.08\times10^{-4}$ & $2.25\times10^{-4}$ & $1.12\times10^{-4}$
		   & $6.31$ & $-$ & $6.41\times10^{-5}$ & $2.19\times10^{-4}$ \\
F-R15-b4 & $3.86\times10^{-3}$ & $1.32\times10^{-4}$ & $3.32\times10^{-3}$ & $1.09\times10^{-3}$
		   & $3.17$ & $6.27$ & $3.18\times10^{-4}$ & $6.35\times10^{-5}$ \\
F-R15-b5 & $2.63\times10^{-3}$ & $2.58\times10^{-4}$ & $2.91\times10^{-4}$ & $9.25\times10^{-5}$
		   & $5.88$ & $-$ & $7.68\times10^{-5}$ & $9.40\times10^{-5}$ \\

\hline\hline
\end{tabular}
\end{center}
Maxwell and Reynolds stresses $\alpha^{\rm Max}$ and $\alpha^{\rm Rey}$ are
measured in the disk zone ($-z_b<z<z_b$), wind stresses $T_{z\phi}^{\rm Max}$
are averaged values measured at the wind base $z=\pm z_b$. All 3D simulations
eventually have odd-$z$ symmetry. Also note that for simulations with weaker net
vertical field, the Alfv\'en points are not contained in the simulation box.
\end{table*}

In Figure \ref{fig:scl}, we further show the radial velocity profiles across the strong
current layer in the three runs at $1, 3, 5$ AU with fixed $\beta_0=10^5$. We see
that although the thickness of the strong current layers vary considerably in the
three cases, their velocity profiles have very similar shapes, with sharper decrease
to the midplane side (since resistivity increases). Moreover, at 5 AU, we see that
the strong current layer is already very thick, and extends even to the disk
midplane. This indicates that the midplane region is already partially coupled to
the magnetic field, and we find that in this run, the Elsasser numbers at midplane
are $\Lambda\approx Am\approx0.1$. This radius, as we shall discuss in
Section \ref{sec:3dresults}, is close to the outer boundary of the pure laminar accretion
zone .

\subsection[]{Unsteady Runs}

Among all our quasi-1D simulations, we find that three of the runs never settle into
a laminar state (S-R3-b6, S-R5-b6 and S-R8-b5), which are not included in Table
\ref{tab:1dresults}. These runs show obvious signs of time variability
all the time, where all the physical quantities exhibit large amplitude variations
especially in the surface layer of the disk, and mass ejection occurs episodically.
We have further tried to initialize runs S-R3-b6 and S-R5-b6 from the time and
horizontally averaged profiles from their 3D counterparts, while they evolve into
similar situations. We therefore conclude that laminar solutions can not be found for
the combination of parameters in these runs. Based on these results, we note that
toward larger radii, progressively larger net vertical flux is required to obtain
a steady laminar solution.

\section[]{Results: 3D Simulations}\label{sec:3dresults}

Obtaining a quasi-1D solution does not necessarily mean that the
solution is stable since non-axisymmetric modes are artificially suppressed.
In paper I, we have demonstrated that at 1 AU, the laminar solution remains
stable in 3D simulations. In this section, we first check whether the laminar
quasi-1D solutions we obtained in the previous section persist in 3D, and we
further explore the gas dynamics in the outer region (up to 15 AU) to identify
the transition from the pure laminar wind to the onset of the MRI.
For all the 3D simulations, we show the standard diagnostic quantities in Table
\ref{tab:3dresults}, and they will be discussed in the following subsections.

\subsection[]{Stability of quasi-1D Results}\label{sec:3dresultslaminar}

We first consider the two 3D runs F-R3-b5 and F-R5-b5 where laminar solutions
were found in quasi-1D simulations. As can be seen in Table \ref{tab:3dresults},
the value of essentially all their diagnostics agree very well with their quasi-1D
counterparts. The 3D runs exhibit very weak level of turbulent Reynolds stress
in the disk zone that is of the order $10^{-6}$ or less in natural unit. This level is
much smaller than the level reported in \citet{ShenStone06} from the decay of
hydrodynamic turbulence, which leaves small amplitude linear modes in
shearing-box whose amplitude is well preserved due to the low-level numerical
dissipation in the Athena code. Therefore, it is consistent with pure laminar flow.
The slight difference in other measured quantities such as $T_{z\phi}^{\rm Max}$
and $\dot{M}_w$ between runs S-R5-b5 (even symmetry) and F-R5-b5 (odd
symmetry) can be attributed to the difference in their symmetry, as one
compares runs S-R3-b5a and S-R3-b5b. In sum, we confirm that the quasi-1D
results discussed in Section \ref{sec:1dresults} are robust in 3D.

\subsection[]{A Fiducial Case}\label{ssec:marginal}

\begin{figure*}
    \centering
    \includegraphics[width=180mm]{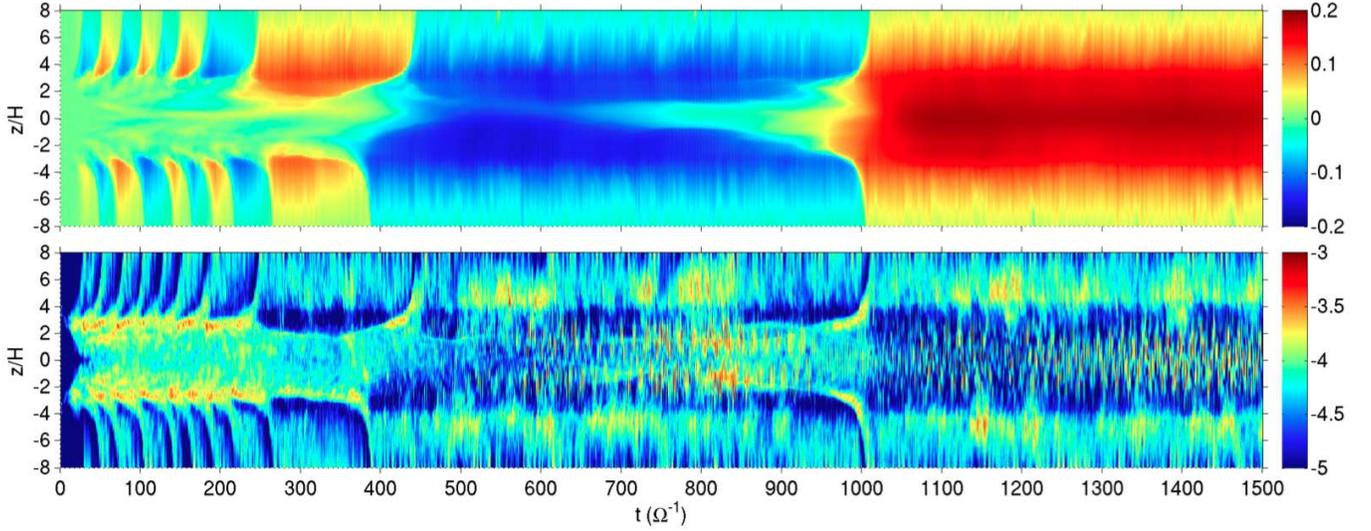}
  \caption{Space-time plot of horizontally averaged toroidal magnetic field
  $\bar{B_y}$ (upper panel) and vertical kinetic energy
  $\log_{10}(\overline{\rho v_z^2/2})$ (lower panel) for our 3D run F-R10-b5.
  The system reaches the final configuration after $t\approx1000\Omega^{-1}$.
  See Section \ref{ssec:marginal} for more details.}\label{fig:fid3dhist}
\end{figure*}

Results presented in Section \ref{sec:1dresults} indicate that laminar solution
becomes impossible at larger disk radii and smaller net vertical magnetic
flux. The boundary resides at 3-5 AU for $\beta_0=10^6$ and $\sim8$ AU
for $\beta_0=10^5$. Correspondingly, we conduct 3D runs for these
parameters (except replacing 8 AU to 10 AU). These 3D
runs show qualitatively similar features in their saturated states, and here
we take run F-R10-b5 as a standard example.

\begin{figure*}
    \centering
    \includegraphics[width=170mm]{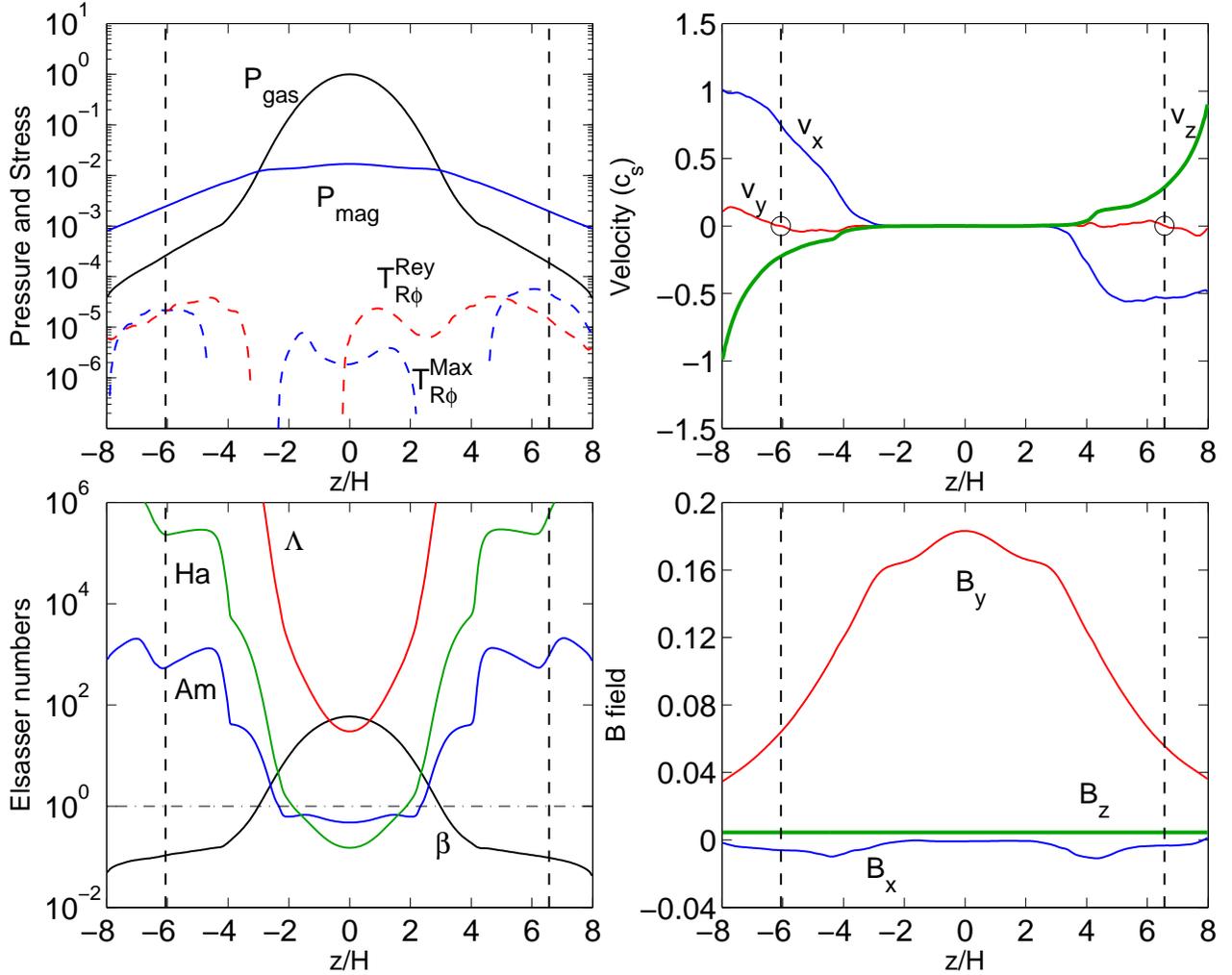}
  \caption{Same as Figure \ref{fig:fid1d}, but for 3D run F-R10-b5 at 10 AU
  with $\beta_0=10^5$. For the upper left panel, we show turbulent (rather
  than the full) contribution to the $R\phi$ component of Maxwell and
  Reynolds stresses. For the upper right panel, the Alfv\'en points are not
  shown since they are not contained in the simulation domain.}\label{fig:fid3dprof}
\end{figure*}

In Figure \ref{fig:fid3dhist} we show the space-time plot of the horizontally
averaged toroidal magnetic field $\overline{B_y}$ and vertical kinetic energy
$\log_{10}(\overline{\rho v_z^2/2})$ for run F-R10-b5. We see that
at the beginning, the system is prone to the MRI and evolves into MRI
turbulence. This is most evident from the characteristic ``Butterfly" pattern
in the first $\sim40$ orbits, where the mean $B_y$ changes sign over a
(quasi-) period of about $10$ orbits\footnote{This dynamo pattern is most
well-known in ideal MHD simulations with zero net vertical magnetic flux, while
it is also present in the presence of net vertical flux but becomes progressively
less periodic with increasing net flux \citep{BaiStone13a}. The periodicity of the
dynamo pattern also changes in the presence of AD \citep{Simon_etal13a}.}.
In this initial phase, the mean toroidal field at disk midplane is very weak
(comparable to net vertical field), a field geometry that is favorable for the
MRI in the AD dominated regime (at disk midplane). We find that the MRI
turbulence is mainly driven from the midplane region, giving rise to stress
level of $\alpha_{\rm turb}\lesssim10^{-3}$ and is responsible for the dynamo
behavior. The base of the FUV layer is also MRI-unstable, but its activity is
overwhelmed by the magnetic dynamo behavior buoyantly rising from the
midplane. Subsequently, the dynamo disappears,
accompanied by the amplification of the mean $B_y$ near the midplane, yet
the mean $B_y$ still undergoes a local minimum across the disk midplane.
The system appears to be adjusting itself between $t=200-1000\Omega^{-1}$,
arriving at the final steady-state configuration from $t=1000\Omega^{-1}$
on, where the mean $B_y$ profile has a single peak at the midplane.
It appears that the strength of the midplane mean $B_y$ largely controls the
turbulent activities: strong $\overline{B_y}$ leads to weaker midplane
turbulence, as one compares the two panels of Figure \ref{fig:fid3dhist}.
It is unclear what controls the mean toroidal field around disk midplane, but
the fact that it takes the system more than 150 orbits to settle to the final state
is indicative of its marginal nature on the transition from purely laminar flow to
the onset of the MRI. To obtain reliable time-averaged quantities, we have run
the simulation further to $t=1500\Omega^{-1}$.

Now we focus on the final saturated state of the simulation since
$t=1080\Omega^{-1}$.
From the space-time plot we see large time variabilities on $\rho v_z^2$
take place at both midplane and the surface FUV layer, indicating turbulence
is driven at both locations. The magnetic field is dominantly toroidal, and
magnetic fluctuations due to turbulence are much weaker compared with the
mean toroidal field, allowing us to extract time-averaged vertical profiles. In
Figure \ref{fig:fid3dprof}, we show the time-averaged vertical profiles of
various quantities analogous to Figure \ref{fig:fid1d}. The only difference is
that the dashed curves on the top left panel correspond to the turbulent
component of (rather than the full) $T_{r\phi}$.
We observe the following features.

1). AD is the dominant non-ideal MHD
effect all over the disk. It is strongest at disk midplane, and becomes
weaker toward disk surface due to X-ray and FUV ionizations. Ohmic
resistivity is totally negligible (with $\Lambda\gtrsim50$).

2). Both the midplane region and the surface FUV layer show enhanced
(but still weak) turbulent stress, and turbulence is largely hydrodynamical
($\alpha_{r\phi, {\rm turb}}^{\rm Rey}>\alpha_{r\phi, {\rm turb}}^{\rm Max}$).
The turbulence at the FUV layer is much more vigorous since the density
is much smaller compared with the midplane region.

3). An outflow is launched, but the flow velocity is much slower compared
with the laminar case. The Alfv\'en point is not contained with the
simulation box, hence the properties of the outflow may be affected more
by the vertical boundary condition (e.g., the feature of the $v_y$ profile near
vertical boundary in Figure \ref{fig:fid3dprof}). We note that the Alfv\'en point
moves away from midplane as one reduces the net vertical magnetic flux
(\citealp{BaiStone13a}, paper I), the fact that the Alfv\'en point falls beyond
the simulation box is consistent with smaller net magnetic flux and weaker
outflow.

The MRI turbulence in the midplane region merits some more discussion.
Within $|z|<2H$, we see from Figure \ref{fig:fid3dprof} that $Am\sim0.5$
with a flat vertical profile, and $B_y\sim0.18$ in code unit with a relatively
flat vertical profile as well. We note that the presence of mean toroidal
field significantly changes the dispersion relation of the MRI. Using the
general dispersion relation derived by \citet{KunzBalbus04}, we find that
given these parameters, the most unstable MRI mode correspond to
$2\pi/k_z\approx H-3H$, with fastest growth rate of $0.018\Omega^{-1}$,
and $k_x\approx k_z$.
Therefore, the most unstable modes can be
properly fitted into our simulation box. According to the simulation results
from \citet{BaiStone11}, under the most favorable magnetic geometry,
the resulting turbulent stress is of the order $4\times10^{-3}$ for
$Am\sim0.5$ (see their Equations (24) and (26)). In our case, the field
geometry is highly toroidal field dominated with
$\overline{B_y}/\overline{B_z}\approx40$.
Note that this is close to the least favorable field geometry for $Am$ of
order unity \citep{BaiStone11}, and the low level of turbulent stress
($\lesssim10^{-4}$, see Table \ref{tab:3dresults}) in the midplane is
consistent with theoretical expectations\footnote{Here we
are comparing the turbulent stress (rather than total stress) with the
the unstratified simulation results of \citet{BaiStone11} because the
stress due to large-scale magnetic fields in unstratified shearing-box
simulations is zero.}.

In sum, we find that in this marginal case, Ohmic resistivity becomes
irrelevant and MRI operates at both disk midplane and surface
FUV layer. The level of the turbulence and efficiency of turbulent
angular momentum transport in the midplane is low because of a
toroidal-dominated magnetic field configuration near the midplane
region. The surface FUV layer is much more turbulent but is not
very efficient to transport angular momentum due to its low density.


\subsection[]{Parameter Dependence}

\begin{figure}
    \centering
    \includegraphics[width=85mm]{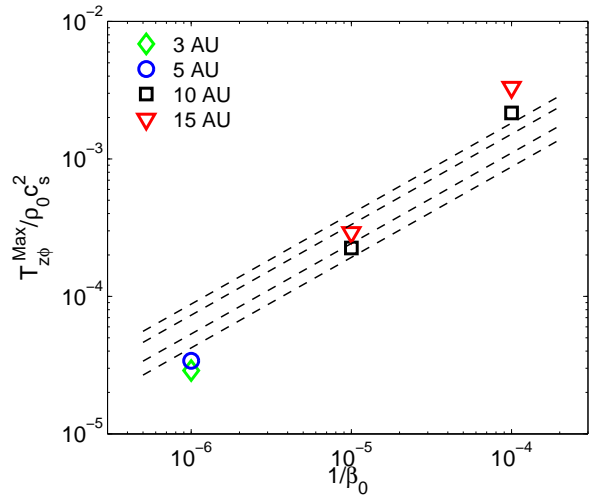}
  \caption{The wind stress $T_{z\phi}^{\rm Max}$ as a function of net vertical
  magnetic flux (characterized by $\beta_0$) measured from all our 3D
  simulations with midplane turbulence. Different symbols correspond to
  simulations at different radii. Dashed lines show the fitting formula
  (\ref{eq:fitTm}) obtained from (quasi-1D) laminar wind simulations for each
  radii (from 3AU to 15 AU, bottom to top). All quantities are in code units.
  Note significant deviations from the laminar-wind fitting formula.}\label{fig:3dparam}
\end{figure}

In this subsection we discuss the rest of the 3D simulations (all except
the two mentioned in Section \ref{sec:3dresultslaminar}).
We start from the two runs at 3 AU and 5 AU with $\beta_0=10^6$. The
midplane region in these two cases still has large Ohmic resistivity, hence
the MRI does not operate there. However, very weak MRI turbulence is
excited in the FUV layer. From Table \ref{tab:3dresults}, we see that the
turbulent stresses are systematically larger than the two 3D companion
runs with $\beta_0=10^5$ (i.e., laminar wind cases), indicating that MRI
is in operation. This is further confirmed by looking at the MRI dispersion
relation. At the FUV front (located at $z\sim\pm4H$), the gas density in
these two cases is around $5\times10^{-4}$. With $\beta_0=10^6$, the
most unstable wavelength is approximately $0.5H$. Although this simple
analysis ignores stratification, the relatively small wavelength indicates
the viability of the MRI.
However, due to the low density in the FUV layer, the resulting value
of $\alpha_{\rm turb}$ is extremely small ($\sim10^{-5}$).

For the four runs at 10 AU and 15 AU, AD becomes the dominant non-ideal
MHD effects throughout the disk, and MRI turbulence sets in both from the
midplane and from the FUV layer. The properties of these runs at saturation
are all very similar to the fiducial case F-R10-b5 discussed in detail earlier.
Among them, run F-R10-b4 appears closest to the laminar case, with a
negative turbulent Maxwell stress ($\alpha_{\rm turb}^{\rm Max}$) and also a
smallest $\alpha_{\rm turb}^{\rm Rey}$ compared with the other three. This
run is parametrically similar to our quasi-1D run S-R8-b4 where pure laminar
configuration is stable. Being located slightly outward, it does show some
turbulent behavior and we find that both the midplane region and the FUV layer
is marginally unstable to the MRI. At this location of 10 AU, we expect a purely
laminar configuration with slightly stronger net vertical flux. Similarly, we expect
unambiguous MRI turbulence at larger radii with fixed $\beta_0$, which is
indeed the case as in our run F-R15-b4.


For 3D runs with MRI turbulence, we find that the scaling relations
(\ref{eq:fitmw}) and (\ref{eq:fitTm}) no longer hold. As an example, we
show in Figure \ref{fig:3dparam} the wind stresses from these 3D runs
and compare them with expectations from the laminar wind fitting formula.
We find that in the presence of turbulence, the dependence of wind stress
on net vertical magnetic flux is more steep: $T_{z\phi}^{\rm Max}$ from
runs F-R3-b6 and F-R5-b6 fall significantly from expectations, while
$T_{z\phi}^{\rm Max}$ from runs F-R10-b4 and F-R15-b4 are well above
(but note the caveat that in simulations with weaker fields, the Alfv\'en
points are not contained in the simulation box).
These results set the radial range where the laminar wind fitting formulas
are applicable.

We note that in all these 3D simulations, turbulence only contributes to a
small fraction of the total stress $T_{r\phi}$. The dominant contribution is
from the Maxwell stress due to large-scale magnetic field (i.e., magnetic
breaking). Using Equation (\ref{eq:accrete}), we see that to have an
accretion rate of $10^{-8}M_{\bigodot}$ yr$^{-1}$, the required value of
$\alpha$ is about $4-5\times10^{-3}$ at $10-15$ AU. From the simulations,
the measured $\alpha^{\rm Max}$ is a bit small but close to this level.
We also note that in quasi-1D runs at inner disk radii, the level of
$\alpha_{\rm Max}$ at $\beta_0=10^4$ is also substantial. Also from 1D
simulations, we see that under the physical symmetry (with field flip and a
strong current layer), the value of $\alpha_{\rm Max}$ is slightly smaller
than (but on the same order of) the case under the unphysical odd-$z$
symmetry. These results together suggest magnetic breaking can
potentially play a non-negligible role in angular momentum transport
\footnote{Most likely only at $R\gtrsim10$ AU. In the inner disk,
$\beta_0\gtrsim10^5$ is sufficient for rapid wind-driven accretion, where
contribution from $\alpha_{\rm Max}$ is less than $\sim10\%$.}.

For a physical scenario where horizontal field flips within the disk, our
quasi-1D runs have implied that at larger disk radii, the flip is likely to
take place close to the midplane at large radii, and the thickness of this
``strong current layer" can be large ($\gtrsim H$). If this is the case, the
midplane region is likely to have very weak toroidal field, and such field
geometry is more favorable for the MRI turbulence in the AD dominated
regime. It is likely that radial transport of angular momentum (i.e., MRI
and magnetic breaking) becomes progressively important toward the
outer disk, while at least up to the range of $10-15$ AU, magnetocentrifugal
wind can still play a dominant role in this process: the required level of
wind stress $T_{z\phi}^{\rm Max}$ at $10-15$ AU is about
$1.5-2\times10^{-4}\rho_cc_s^2$ (for accretion rate of
$10^{-8}M_{\bigodot}$ yr$^{-1}$), which is easily accounted for with
$\beta_0=10^5$.


\section[]{Summary and Discussions}\label{sec:discussion}

\subsection[]{Criteria for the Laminar Wind Solution}\label{ssec:criteria}

\begin{figure}
    \centering
    \includegraphics[width=90mm]{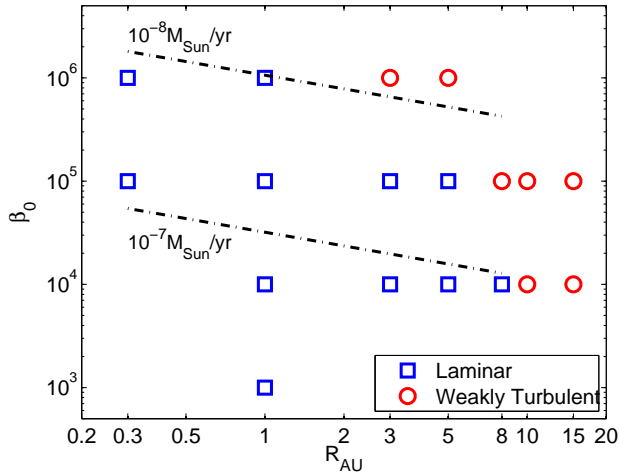}
  \caption{Summary of the parameter study from simulations in paper I and this
  paper. Each symbol denote one particular set of parameter in the
  $R_{\rm AU}-\beta_0$ plane, assuming a MMSN disk model. Blue squares
  denote the existence of laminar wind solution, while red circles mean the disk
  has to be turbulent, though turbulence is very weak with stress level of the order
  $10^{-4}$ or less. Two dash-dotted lines indicate the desired $\beta_0$ as
  a function of radius for wind-driven accretion rate of $10^{-8}$ and
  $10^{-7}M_{\bigodot}$ yr$^{-1}$ (based on Equations (\ref{eq:accrete}) and
  (\ref{eq:fitTm}) for MMSN disks).}\label{fig:summary}
\end{figure}

Piecing together all simulations presented in the present paper and paper I,
we show in Figure \ref{fig:summary} the parameter space we have explored,
indicating whether a laminar solution is possible. Clearly, laminar solutions can
be found in the inner disk less than $10$ AU. The radius where turbulence sets
in depends on the strength of the net vertical magnetic field. With relatively
strong net vertical field $\beta_0\sim10^4$, the purely laminar zone could extend
to about 10 AU, while for much weaker net vertical field $\beta_0\sim10^6$, the
purely laminar zone could shrink to 3 AU. We emphasize that even though
we have labeled ``turbulent" in much of parameter space in Figure \ref{fig:summary},
the level of turbulence can be extremely weak. In particular, for the two symbols at
3 AU and 5AU with $\beta_0=10^6$, extremely weak turbulence only exists in the
the surface FUV layer, and the bulk of the disk is still largely laminar.
Even at 10 AU and 15 AU, where turbulence exists at both disk midplane and
the FUV layer for $\beta_0\gtrsim10^4$, the transport coefficient is of the order
$\alpha\sim10^{-4}$ or less.

Using Equations (\ref{eq:accrete}) and (\ref{eq:fitTm}) and under the assumption
that the laminar solution exists, we further draw two dash-dotted lines in Figure
\ref{fig:summary} showing the expected parameter for the disk to maintain
steady state wind-driven accretion rate of $10^{-8}$ and $10^{-7}M_{\bigodot}$ yr$^{-1}$.
It clearly shows the importance of the laminar wind solutions at the inner disks
within 10 AU. Although the parameter space marked by red circles does not
obey our fitting formula (\ref{eq:fitTm}), the deviation is not substantial (see
Figure \ref{fig:3dparam}), and we may still use the dash-dotted lines as a reference.

Combining the results with the studies in paper I, the criteria for the existence of
the laminar wind solution can be readily inferred. We emphasize that we do {\it not}
attempt to provide quantitative and general criteria for MRI suppression and wind
launching. Instead, we design these criteria specifically for qualitatively
understanding the launching laminar winds from PPDs, as follows.
\begin{itemize}
\item Strong Ohmic resistivity ($\Lambda<1$) in the midplane region,
which is necessary to suppress the MRI around disk midplane.

\item Strong AD dominated disk upper layer, which is essential to suppress the
MRI at disk surface.

\item The presence of (not too weak) net vertical magnetic flux, otherwise the
MRI does operate but works extremely inefficiently.

\item Sufficient ionization beyond disk surface ($Am\gtrsim100$), which is
essential for wind mass loading.
\end{itemize} 

We note that the dominance of AD in the disk surface layer is always the case
due to the low density, while strong Ohmic resistivity at disk midplane is only
possible in the inner region of PPDs. This is the essence why such laminar wind
solution does not extend to the outer disk beyond 10 AU as we have confirmed
in this paper. The requirement of not-too-weak magnetic field is mainly to avoid
the MRI in the surface FUV layer. Since the FUV ionization penetrates deeper
in the outer disk (under the assumption of constant penetration column),
stronger field (measured in $\beta_0$) is required toward outer disk to suppress
the MRI in the FUV layer, which explains the trend in Figure \ref{fig:summary}.


Another point of clarification again involves the magnetic field strength.
It is well known that the MRI can be suppressed with sufficiently strong
magnetic field with/without non-ideal MHD effects. In fact, conventional
wind launching criterion which requires near-equipartition field strength is
partly to avoid the MRI. In PPDs, if field strength approaches this level,
other criteria outlined in this subsection would become meaningless. However,
the conventional wind scenario would result in excessively strong mass loss
and accretion unless the disk is substantially depleted \citep{CombetFerreira08}.
Also, wind launching will be dramatically suppressed for field strength beyond
equipartition \citep{Shu_etal08,Ogilvie12}. Therefore, throughout this paper, we
always limit ourselves to field strength much weaker than equipartition
($\beta_0\gg1$). The main reason we are able to find laminar wind solution
with field strength much weaker than equipartition is that we have adopted a
realistic ionization profile (rather than constant Elsasser number profile), where
suppression of the MRI by Ohmic dominated midplane is essential.



\subsection[]{Magnetic Flux Distribution for Steady-State Accretion}\label{ssec:accretion}

In paper I, we have emphasized the importance of net vertical magnetic flux: it is
the sufficient and necessary requirement to drive rapid accretion in the inner
region of PPDs. In the outer region of PPDs (e.g., beyond 30 AU) where MRI-driven
accretion is likely to dominate \citep{Bai11b}, the presence of net magnetic flux has
also been shown to be crucial \citep{Simon_etal13a,Simon_etal13b}, otherwise the
MRI would simply be too inefficient. The rate of angular momentum transport due to
either disk wind or the MRI depends sensitively on the amount of net vertical magnetic
flux threading the disk. Therefore, the global distribution of poloidal magnetic flux in
PPDs is the key to understanding the accretion process in PPDs.

Here we focus on the inner disk region with laminar accretion, and ask the question
of what distribution of poloidal magnetic flux would result in steady state accretion
with accretion rate consistent with observations? Our starting point is the fitting
formula (\ref{eq:fitTm}) in Section \ref{ssec:1dresultsparam}. Following the discussion in
that section, we convert $\beta_0$ in the MMSN disk model to physical magnetic
field strength,
and further use Equation (\ref{eq:accrete}), to find the wind-driven accretion rate to be
\begin{equation}
\dot{M}\approx0.91\times10^{-8}M_{\bigodot}\ {\rm yr}^{-1}{R_{\rm AU}}^{1.21}
\bigg(\frac{B_{z0}}{10\ {\rm mG}}\bigg)^{0.93}\ .
\end{equation}
We note that this formula applies only in the laminar region of the disk.
Uncertainties with this formula mainly arise from the treatment FUV ionization,
and the assumption of the MMSN temperature profile, but they should only
introduce minor corrections (recall from Table 4 of paper I, varying the FUV
penetration depth by a factor of 10 leads to difference in $T_{z\phi}^{\rm Max}$
by a factor of about 2).


Clearly, this formula shows that large-scale poloidal field strength of about
$10$ mG at 1 AU is needed to drive accretion with the typical observed rate
of $10^{-8}M_{\bigodot}$ yr$^{-1}$. We emphasize that surface density does
not enter the formula above, a result identified in paper I, and sustaining
steady-state accretion requires that the radial distribution of net vertical
magnetic field follows $B_z(R)\propto R^{-1.3}$. In other words, using
$\Phi_p(R)$ to denote the total poloidal magnetic flux contained within disk
radius $R$, it should follow the law of $\Phi_p(R)\propto R^{0.7}$.

Questions remain regarding the transport of poloidal magnetic flux in
PPDs: how can poloidal magnetic flux be arranged to arrive at the desired
configuration. Most studies on magnetic flux transport in accretion disks
assume balance between advection due to viscous accretion and diffusion
due to resistivity \citep{Lubow_etal94a,GuiletOgilvie12}. This is not the
case for wind-driven accretion. As was discussed in paper I, the solutions
obtained in our simulations has zero toroidal electric field, corresponding
to a stationary magnetic flux distribution. In other words, inward advection
due to wind-driven accretion is balanced exactly by the outward diffusion
in the strong current layer. Although this is mainly a consequence of the
shearing-box approach, such stationary magnetic flux distribution is
desirable property of the wind solution. We note that other wind solutions
are still possible with non-zero advection velocity of magnetic flux, which
generally require global approach.

\subsection[]{Global Picture}\label{ssec:global}

\begin{figure*}
    \centering
    \includegraphics[width=170mm]{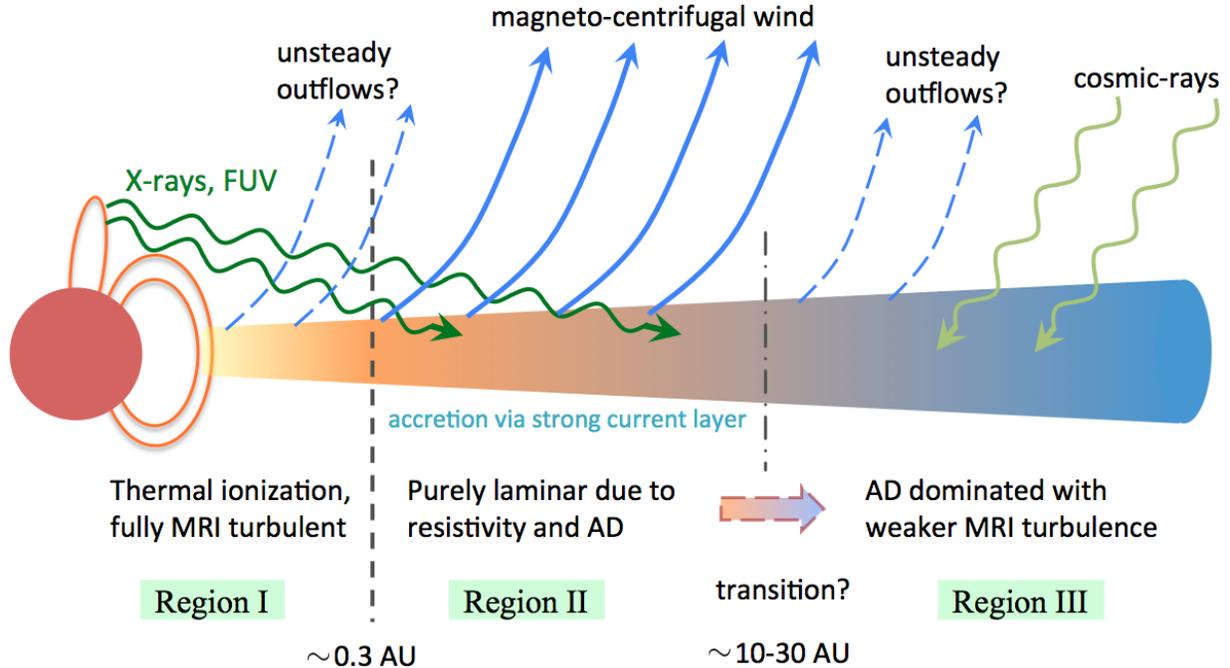}
  \caption{Schematic global picture of protoplanetary disks (not to the scale)
  in light of our new simulation results that incorporate the effect of AD. See
  Section \ref{ssec:global} for detailed discussions.}\label{fig:newpic}
\end{figure*}

Simulations from our paper I demonstrated a robust result that the MRI is likely
to be suppressed in the inner region of PPDs, while magnetocentrifugal wind
can be launched due to the large-scale poloidal magnetic flux threading the
disk. Compiling further from all simulations presented in this paper, which cover
a much wider range of disk radii, a new global picture on the accretion process
in PPDs emerges. This new scenario is illustrated schematically in Figure
\ref{fig:newpic}, which contrasts with the conventional picture of layered
accretion (\citealp{Gammie96}, see also Figure 7 of \citealp{Armitage11}).

Overall, the PPD can be divided into three bulk regions.
\begin{itemize}
\item The innermost region (region I), where the gas is sufficiently hot
($T\gtrsim10^3$K) to thermally (collisionally) ionize Alkali metals (Na and K).
The gas behaves in the ideal MHD regime, and the entire disk is highly
turbulent due to the MRI.

\item The inner region (region II), where the disk is largely laminar due to the
combined effects of Ohmic resistivity and AD. Angular momentum transport is
dominantly driven by the magnetocentrifugal wind, and accretion proceeds
through a strong current layer that is offset from disk midplane.

\item The outer region (region III), where MRI is likely to be the dominant
mechanism for angular momentum transport, but its strength is relatively weak
due to strong AD.
\end{itemize}
Below we briefly discuss each region separately, as well as the transition
between them.

In region I, the gas dynamics can be inferred from ideal-MHD shearing-box
simulations of the MRI. We note that since net vertical magnetic flux is
essential for driving rapid accretion for other parts (regions II and III) of the
disks, such net vertical flux is also likely to be present in the innermost region
of the disk. Moreover, since this region is close to the star, it can also be
threaded with some stellar magnetic flux. The behavior of the MRI in the
presence of net vertical flux in the ideal MHD regime has recently been
explored in great detail
\citep{SuzukiInutsuka09,Fromang_etal12,BaiStone13a}.
It was found that an outflow is always launched, which is highly unsteady
and its mass outflow rate increases with net vertical flux similar to our
laminar wind case. However, it was argued that the outflow launched from
an MRI-turbulent disk is unlikely to be {\it directly} connected to a global wind
either due to the MRI dynamo or symmetry
issues \citep{BaiStone13a}. Moreover, strong interaction with the protostellar
magnetosphere is also expected in the inner part of region I, which also
likely launches powerful and highly dynamic outflows \citep{Romanova_etal09}.
Therefore, we use dashed arrows in Figure \ref{fig:newpic} to illustrate such
unsteady outflow whose fate remains to be explored. 

The outer boundary of region I lies where the temperature drops below
about $10^3$ K. Within region I, viscous heating is generally the
dominant heating source, and for typical accretion rate of
$10^{-8}M_{\bigodot}$ yr$^{-1}$, the location where $T$ drops below
$10^3$ K is around $0.3$ AU (e.g.,
Figure 2 of \citealp{Armitage11}, yet the exact location depends on
model parameters). Since thermal ionization depends extremely
sensitively on temperature near the threshold, the transition from
region I to region II should be abrupt, yet it would be very interesting
to explore the gas dynamics near the interface connecting the
fully turbulent region and the full laminar region. 

Region II has been discussed extensively in paper I and this paper. It
covers the radial range where the four criteria identified in Section
\ref{ssec:criteria} are met, and extends from the outer boundary of
region I to some outer radius.
We note that at $3-5$ AU, the disk wind is sufficient to drive rapid
accretion at $3-5$ AU even with weak field $\beta_0=10^6$. Although
the thin FUV surface layer is expected to be weakly turbulent, the bulk
of the disk can still considered to be largely laminar. Therefore, we
may relax our criterion 3 (of Section 5.1) and consider region II to
extend to an outer radius of somewhere between 5-10 AU.

Slightly beyond the outer boundary of region II where Ohmic resistivity
becomes unimportant, MRI starts to operate at disk midplane. However,
because of strong AD and relatively strong
toroidal-domianted magnetic field configuration near midplane, turbulence
resulting from the MRI is very weak. While magnetic breaking due to
large-scale radial and toroidal fields could play a non-negligible role on
angular momentum transport, magnetocentrifugal wind is still most likely
the dominant mechanism to drive disk accretion, provided an even-$z$
symmetry continues to apply. The wind stress and mass loss rate no
longer follow the scaling relations (Equations (\ref{eq:fitmw}) and
(\ref{eq:fitTm})) identified for the laminar wind region (region II) due to
the presence of MRI turbulence at disk midplane and FUV layer.

Region III is not covered in our numerical simulations. However, earlier
works suggest that MRI alone is capable of driving rapid accretion to
feed the inner disk if under a favorable magnetic field geometry and
with the assistance of tiny grains and/or FUV ionization
\citep{Bai11b,PerezBeckerChiang11b}. Here, favorable field geometry
refers to the magnetic configuration with both net vertical and mean
toroidal magnetic fields, with the toroidal component a few times stronger
\citep{BaiStone11}. Recently, \citet{Simon_etal13a,Simon_etal13b}
explored the gas dynamics in region III with shearing-box simulations.
They focused at two disk radii, 30 AU and 100 AU, and found that 1) net
vertical magnetic flux is essential to drive rapid accretion, 2) the MRI-driven
accretion rate is a strong function of net vertical magnetic flux. There is
also outflow from the simulations as a natural consequence of net vertical
flux. While the fate of the outflow is again uncertain for reasons similar to
that in region I, MRI is the dominant source of angular momentum transport.

It is very likely that there is a transition zone that covers a relatively wide
range of radii between region II and region III (e.g., from $\sim5-10$ AU to
$\sim30$ AU), where the dominant angular momentum transport mechanism
shifts {\it gradually} from disk wind to the MRI/magnetic braking. Moreover,
the Hall effect, which is not included in our studies, is the dominant non-ideal
MHD effect in such transition radii. It will be extremely interesting to explore
the transition using global simulations.

\subsection[]{Concluding Remarks}\label{ssec:conclusion}

Our simulations have demonstrated the magnetocentrifugal wind is an
essential ingredient for driving accretion in the inner region of PPDs.
Although we are unable to provide large-scale kinematics of the disk wind
based on our local shearing-box simulations, the range of radii where we
expect such disk wind is consistent with the inferred launching radii of the
low-velocity component of the molecular outflows from observations \citep{Anderson_etal03,Coffey_etal04,Coffey_etal07}. The laminar nature
of the gas flow in the inner disk has important implications on grain growth
in PPDs, planetesimal formation, and potentially planet migration, which
all merit future investigations.

The global picture proposed in this paper represents a large-scale
framework, while the details of the picture remain to be explored with a
lot more efforts. In particular, the Hall effect that was not included in our
simulations is the dominant non-ideal MHD effect in a large portion of
parameter space in PPDs and may play an important role on the MHD
stability of the disk. Moreover, global simulations are essential to
address the stability of the strong current layer, the transition between
regions I, II and III outlined in the previous subsection, and the transport
of large-scale poloidal magnetic flux. Global simulations are also
necessary to quantify the kinematics of the disk wind, such as the rate
and velocity of the mass outflow, and to make direct comparisons with
observations.

\acknowledgments

I thank the referee for useful comments and suggestions that improve the
clarity of this paper.
This work is supported for program number HST-HF-51301.01-A provided by
NASA through a Hubble Fellowship grant from the Space Telescope Science
Institute awarded to XN.B, which is operated by the Association of Universities
for Research in Astronomy, Incorporated, under NASA contract NAS5-26555.


\bibliographystyle{apj}

\label{lastpage}
\end{document}